# The Test of LLRF control system on superconducting cavity

ZHU Zhenglong (朱正龙)[1,3] WANG Xianwu(王贤武)[1] WEN Lianghua(文良华)[4] CHANG Wei(常玮)[1,] ZHANG Ruifeng(张瑞峰)[1] GAO Zheng(高郑)[1,2] CHEN Qi(陈奇)[1,2]

(1 Institute of Modern Physics, Chinese Academy of Sciences, Lanzhou 730000, China;
2 Graduate University of Chinese Academy of Sciences, Beijing 100049, China;
3 Northwest Normal University，Lanzhou 730070, China;
4 School of Physics and Electronic Engineering, Yibin University, Yibin 644000, China)

**Abstract:**
The first generation Low-Level radio frequency(LLRF) control system independently developed by IMPCAS, the operating frequency is 162.5MHz for China ADS, which consists of superconducting cavity amplitude stability control, phase stability control and the cavity resonance frequency control. The LLRF control system is based on four samples IQ quadrature demodulation technique consisting an all-digital closed-loop feedback control. This paper completed the first generation of ADS LLRF control system in the low-temperature superconducting cavities LLRF stability and performance online tests. Through testing, to verify the performance of LLRF control system, to analysis on emerging issues, and in accordance with the experimental data, to summarize LLRF control system performance to accumulate experience for the future control of superconducting cavities.

## 1. Introduction

Accelerator-driven subcritical system (ADS), is an effective way to reduce the radioactivity of nuclear waste [1]. Institute of Modern Physics bear ADS linac RF system operating frequency is 162.5MHz; its linear superconducting cavity using a half-wavelength resonator (HWR) form, in the superconducting conditions the cavity $Q_{Load}$ (Load quality factor) is $1\times10^6$ [2]. Requires the amplitude stability is less than$\pm6\times10^{-3}$, the phase stability less than $\pm0.7°$, the frequency detuning angle of cavity less than$\pm0.6°$, the operating point of electric field accelerate gradient(Eacc) $\geq$ 4.7MV/m, the surface of the cavity peak electric field(Epk)$\geq$25MV/m. According to the characteristics of superconducting cavities for high $Q_{Load}$ value, the bandwidth is about 200Hz. Development of full-digital LLRF control system IQ-based technology, and tested in room temperature on the mold cavity. The amplitude stability is ±5‰(peak to peak), and the phase stability is ±0.3°(peak to peak). This paper summarizes the LLRF control system on liquid helium cryogenic superconducting cavity testing and analysis, and accumulates relevant technical experience for further optimization of the system.

The system had been tested on the room temperature copper cavities, meet the design specifications and control requirements. At room temperature, Q value of the copper cavity is about 2800, wide bandwidth, and does not feed in large power, at room temperature without multiple electron multiplier (Multipacting[3]) exists, and no other large disturbances, LLRF control system can keep stable for a long time. For low-temperature 4.2K operation, QL value as high as 1

×10⁶, the bandwidth is 230Hz, so when the temperature change, helium pressure fluctuations or vacuum change would lead the cavity center frequency easily shift exceed the narrow bandwidth. In this case, it is very challenging for LLRF control system to control the cavity.

This paper will summarize and analyze the testing of LLRF control system.

## 2. LLRF control system

Institute of Modern Physics, in 2012, completed the design and construction of the first generation of digital LLRF control system, and tested on room temperature cavity. LLRF control system overall structure shown in Figure 1.

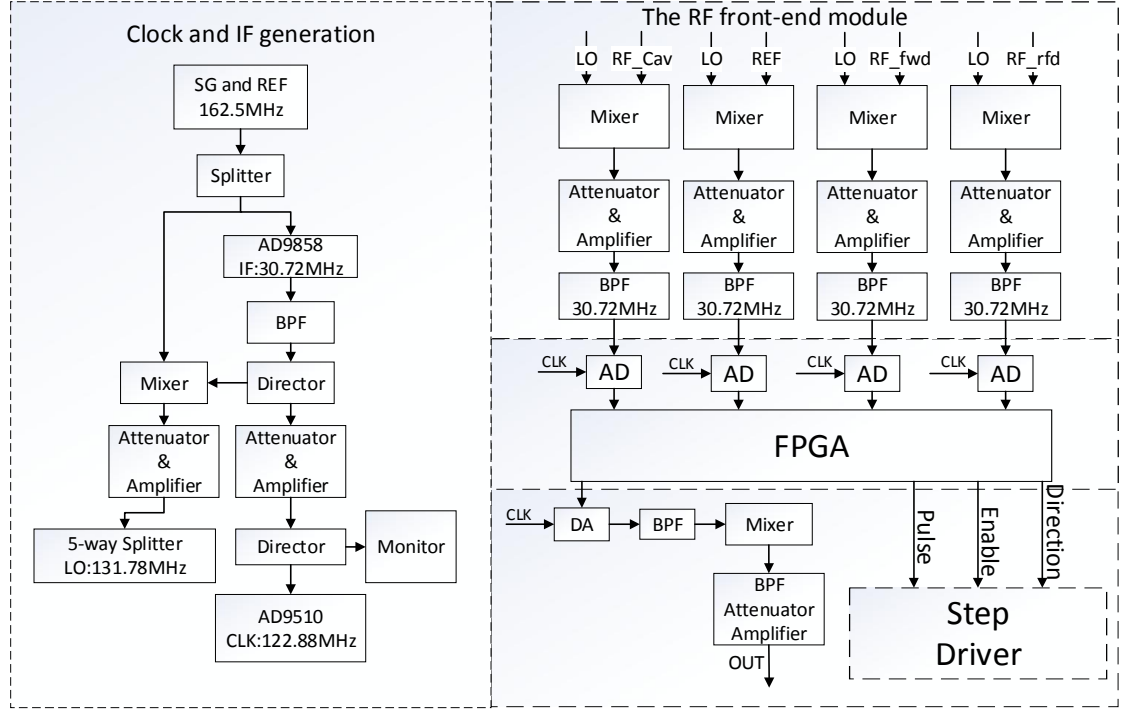

Figure 1: The first generation of digital LLRF block diagram of the overall structure

2.1. signal sampling and recovery

Extracting a signal frequency of the cavity 162.5MHz, down-converted to the 30.72MHz frequency(IF), the sampling clock is 122.88MHz, the sampling frequency is four times the clock frequency of the IF signal, satisfy the IQ sampling theorem. The reason for this is that you can capture directly to the combination sequence IQ, through IQ demodulation can get the I sequence and Q sequence. If we defy the first sample value is I, there is sequence I, Q,-I,-Q, I, Q ......, if IQ sequence represented by S (n), I (n) represents the sequence I, Q (n ) said Q sequences, there are:

$$S(n)=I、Q、-I、-Q、I、Q... \quad (n=1,2...) \qquad (1)$$

$$I(n)=\frac{S(n)-S(n+2)}{2} \qquad (2)$$

$$Q(n)=\frac{S(n+1)-S(n+3)}{2} \qquad (3)$$

## 2.2. PI control

The I and Q sequence by Cordic vector rotation operation, to obtain the amplitude and phase of the RF signal, then the amplitude and phase are respectively controlled by PI and then restore the IF output by the digital DDS. We use Amp (n) and Pha (n) represent the calculated amplitude and phase of Cordic, and use the A and P represent the amplitude and phase set values, use Amp_adjust (n) and Pha_adjust (n) represent the amplitude and phase of the PI controller output, Amp_adjust (n) and Pha_adjust (n) represented as the formula (6), (7) the formula:

$$sum\_a(n) = \sum_{1}^{M}\left(sum\_a(n-1) + Ki*(A - Amp(n))\right), sum\_a(0) = 0 \quad (4)$$

$$sum\_p(n) = \sum_{1}^{M}\left(sum\_p(n-1) + Ki*(P - Pha(n))\right), sum\_p(0) = 0 \quad (5)$$

$$Amp\_adjust(n) = Kp*(A - Amp(n)) + sum\_a(n) \quad n = 1,2\ldots M; \quad (6)$$

$$Pha\_adjust(n) = Kp*(P - Pha(n)) + sum\_p(n) \quad n = 1,2\ldots M; \quad (7)$$

M is the number of samples.

## 3. System test and cavity conditioning

### 3.1. Testing program

LLRF control system consists of two loops, namely the amplitude and phase loop and frequency Tuning loop. In the amplitude-phase loop: the LLRF control system output signal via a variable attenuator control the 20KW power source. Power source output signal, via a directional coupler and then feed into the superconducting cavity through a high power coupler. From the cavity body extracted signal through the amplifier and adjustable attenuator conditioning, through a power splitter divided into two signals, one of signals send into LLRF control system to constitute the amplitude and phase loop. Frequency tuning loop: On the directional coupler to get the Forward signal, and then get the other signal from the power splitter as Reflected signal, detect the phase , then use the algorithm controls the motor drive, complete the tuning loop. LLRF control system test loop components shown in Figure 2.

PC and FPGA communication, through a Gigabit Ethernet. The main function is showing the amplitude and phase information of the cavity, controlling the amplitude and phase loop open or close, and controlling frequency tuning loop open or close. Other sides, PC also configure the AD9858 and AD9510 operating frequency and output signal frequency, output ports.

After completion of the software configuration of each module, adjusting input and output signals properly, adjusting signal source frequency, observing spectrum wave of the extracted signal on the spectrum analyzer, to find the resonant frequency of the cavity. At same time keeping the power source output in a reasonable range. After testing found, at cavity resonance frequency the Multipacting is very serious, we need to implement cavity conditioning for a long time to reduce or eliminate Multipacting.

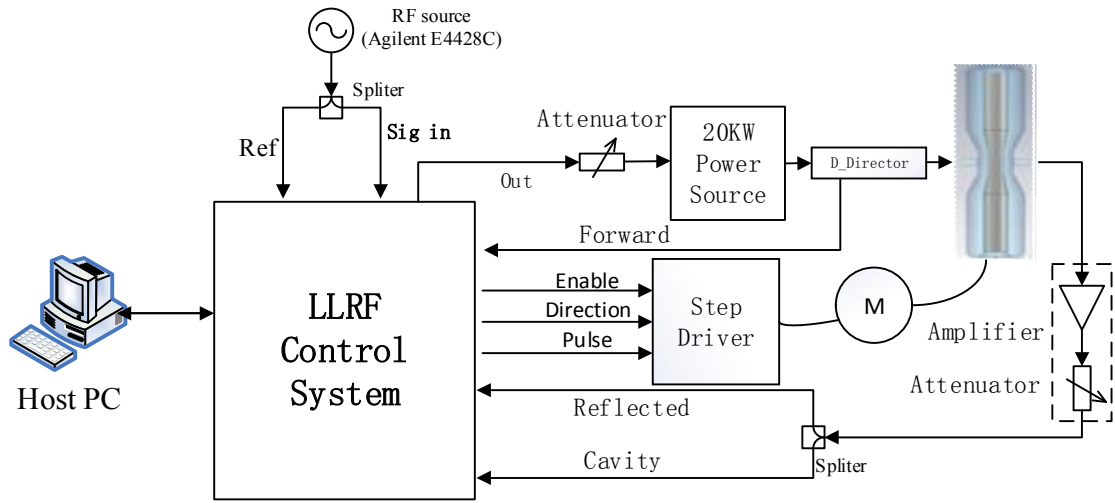

Figure 2: LLRF control system test loop

## 3.2. Cavity conditioning

Start of the test, the superconducting cavity easily aroused Multipacting, the cavity needs to conditioning use pulse add power, the cavity Multipacting scope shown in Figure 4. Using FM-modulated function of Agilent E4428C constitute a VCO-PLL [4] loop to achieve cavity conditioning, tracking resonant frequency of superconducting cavity (Figure 3).

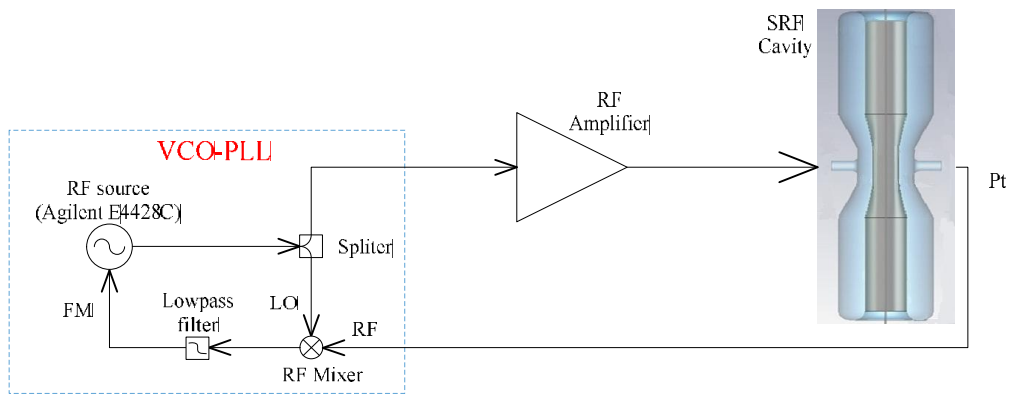

Figure 3: a schematic view of a pulse cavity conditioning

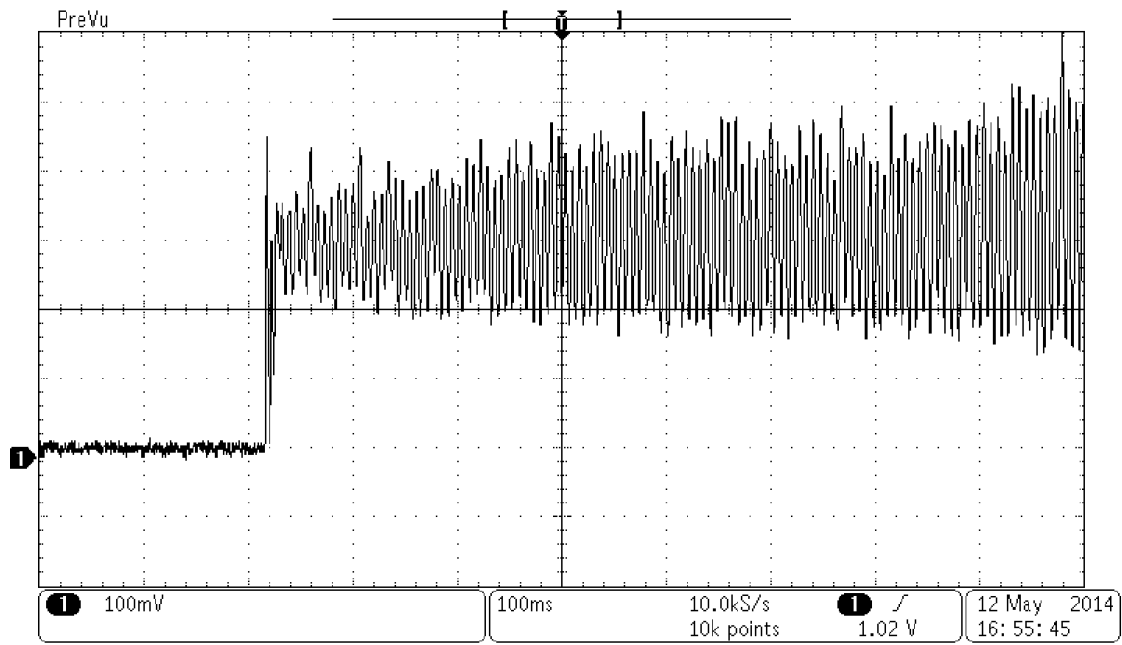

Figure 4: Multipacting via Crystal detector scope

Testing found that when adjusting the frequency of the signal source, so that the cavity frequency deviation resonance frequency, which is not easy to arouse Multipacting, at this condition LLRF control system amplitude-phase loop can be closed. But once we start the closed-loop control system, power source output power abruptly, the cavity along with provoked Multipacting, cavity field disappear. Along with cavity amplitude decreases Multipacting getting small, and then establish a certain strength of the cavity field again, followed Multipacting again, then the process is repeated. In this state, LLRF control system is difficult to control the amplitude and phase stability of the cavity.

Pulse conditioning only training a resonant frequency point Multipacting of the resonant frequency of the cavity. However, the cavity resonance frequency will vary with the helium pressure and temperature change, it is necessary to keep the cavity in a relatively wide band no Multipacting for LLRF control system to normal operation and control. We use Agilent E4428C signal source, set the output signal amplitude is fixed, set the output signal frequency in a certain range and change the frequency of output signal in each step to scan conditioning the cavity.

## 4. Experimental results

After a long time Superconducting cavity scan conditioning, its Multipacting has a certain improvement. LLRF control system is able to control the amplitude and phase of the cavity voltage to remain stable for a long time. Amplitude and phase stability closed-loop of the control interface as shown in Figure 5. The figure shows the signal curve of the amplitude and phase of the cavity, also shows the amplitude and phase loop control interfaces, and the parameter setting of PI controller. Two curves in the figure shows that LLFR control system in closed-loop state keep the amplitude and phase of the cavity signal stable, the closed-loop stable state has continued 45 minutes, Epk reach 25.1MV/m, and can remain stable for a long time.

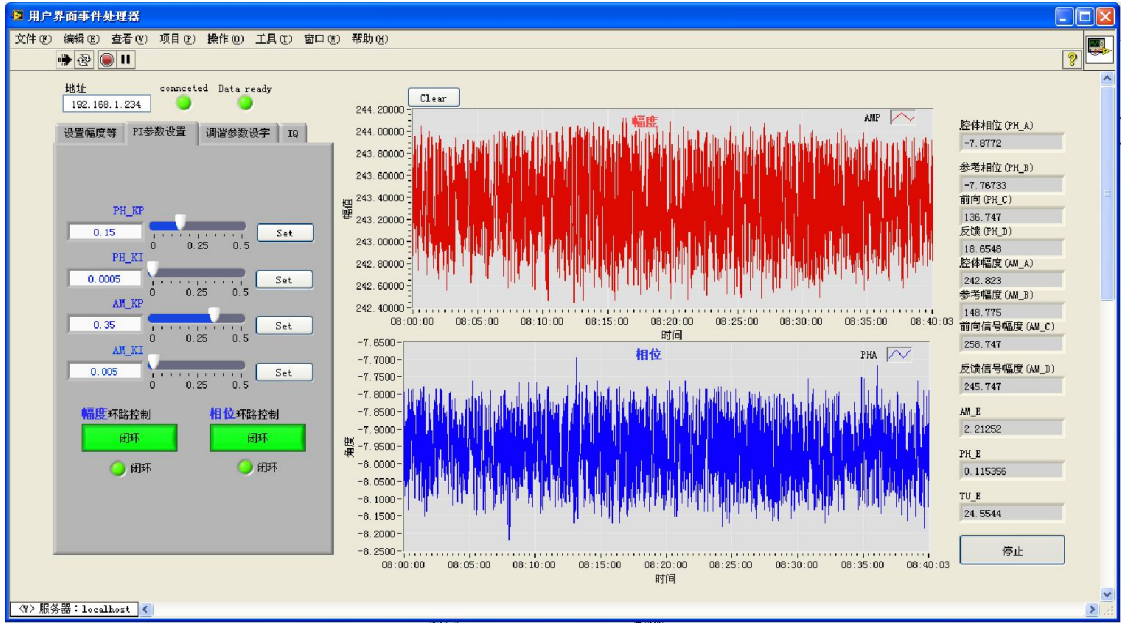

Figure 5: The LLRF control system interface

Figures 6 shows the extraction signal spectrum of the cavity in amplitude-phase open loop state. As can be seen from the figure, in the range of 2kHz Span, cavity resonance peaks within 200Hz. In addition to the main peak, on both sides there are relatively large phase noise. But this cavity Multipacting disappeared, the cavity resonance frequency and the amplitude of the jitter are not large. In this case, it is beneficial for LLRF control system to attempt closed-loop.

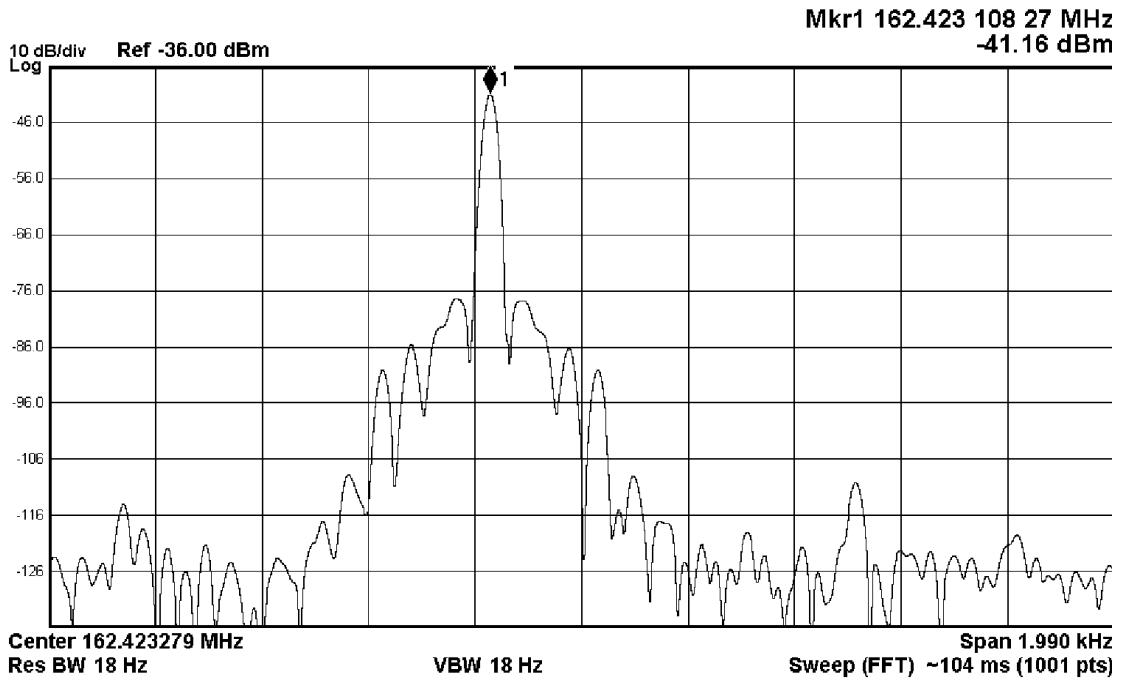

Figure 6: The cavity signal spectrum on open loop

Figure 7 shows the extraction signal spectrum of the cavity in amplitude-phase close loop state. As can be seen from the cavity signal spectrum, in the closed loop condition, the cavity phase noise signal is effectively suppressed. As shown in Figure 7, at this time the cavity signal is relatively stable, the phase noise is suppressed to about -106dBm. However, we also can see from the figure, there are exist two small peaks of phase noise on the both sides of the main peak. This shows that the parameters of PI controller are not optimal. So PI parameters need to be adjusted to suppress the two peaks of the phase noise.

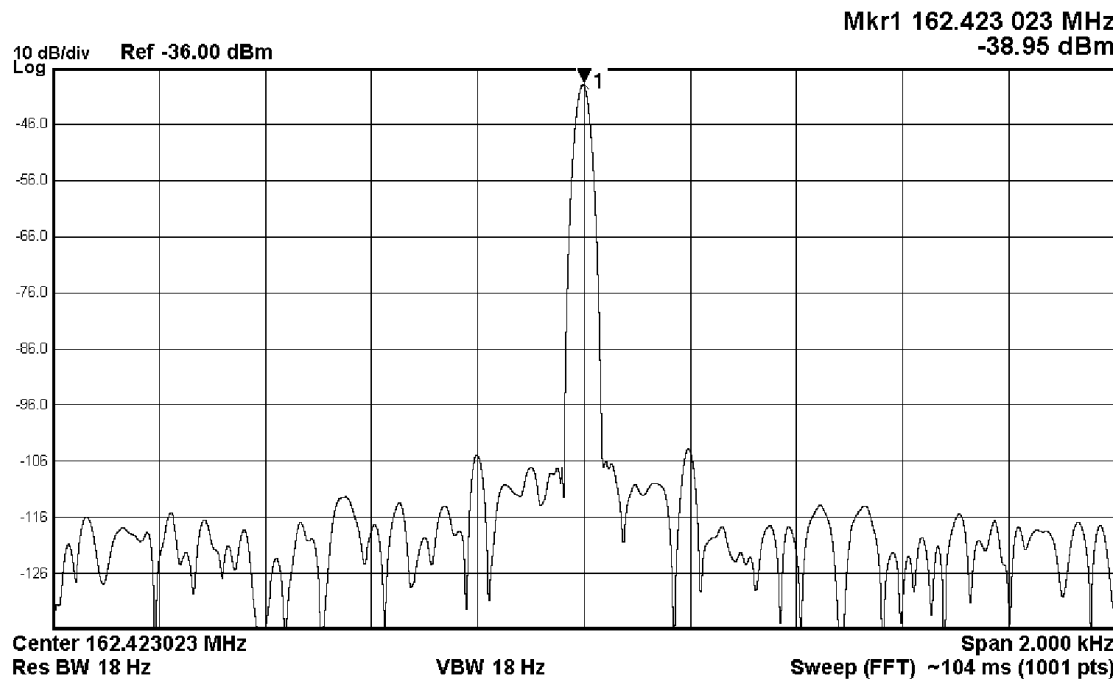

Figure 7: The cavity signal spectrum on close loop

By adjusting the parameters of the PI controller, such as proportional gain Ki. The effect of phase noise suppression of LLRF control system gradually becomes stronger. Figures 7 and 8 are views of cavity signal spectrum of different PI parameters. Shown in Figure 8, the suppressing effect of LLRF control system is very well. In addition to the main resonance peak, other noise signals amplitude of cavity is suppressed very low, and in this state closed-loop state can keep stable for a long time.

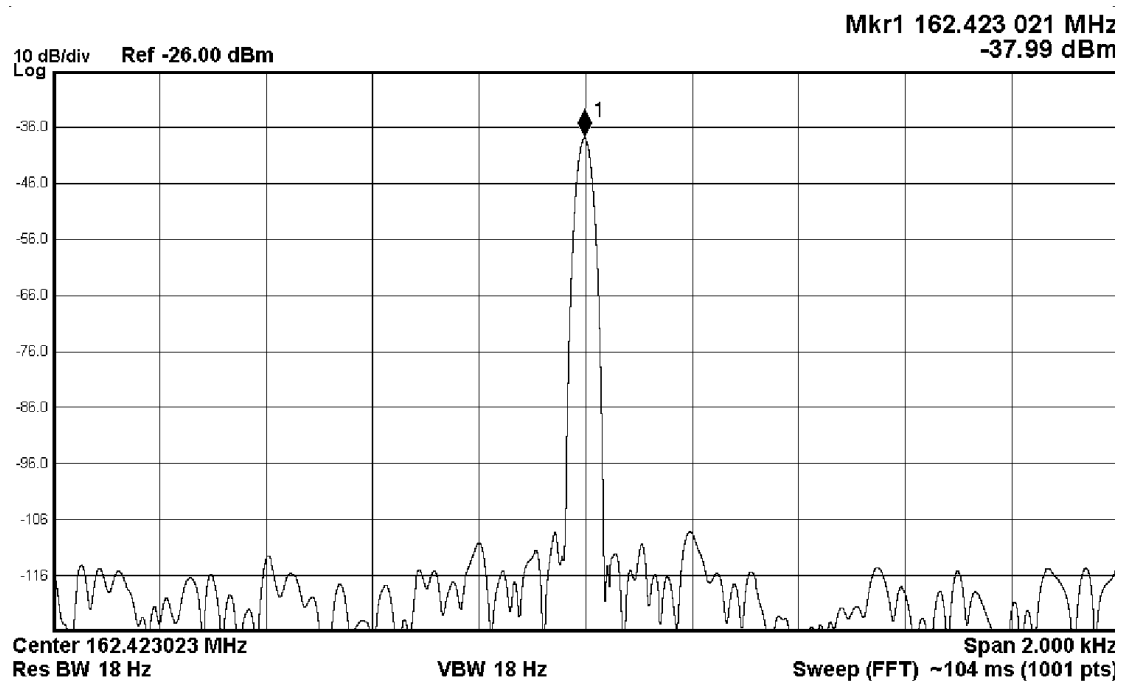

Figure 8: The cavity signal spectrum on close loop

## 5. Conclusion

ADS injector II LLRF control system superconducting cavity test is successful. Superconducting cavity amplitude and phase can be closed-loop controlled stable by LLRF control system, Amplitude stability of ± 3.4 ‰, phase stability of ± 0.3 °, and the cavity Epk reach 25.1MV/m.

The Digital LLRF control system testing on cryogenic superconducting cavities for LLRF control system improvements lays the foundation, Accumulates LLRF system test experience, and lays the foundation for operation online to the LLRF control system.